\documentstyle[aas2pp4]{article}

\def\simless{\mathbin{\lower 3pt\hbox
     {$\rlap{\raise 5pt\hbox{$\char'074$}}\mathchar"7218$}}}   
\def\simmore{\mathbin{\lower 3pt\hbox
     {$\rlap{\raise 5pt\hbox{$\char'076$}}\mathchar"7218$}}}   

\received{}
\accepted{}
\slugcomment{Submitted to Ap.J. Letters, June 17, 1998}

\lefthead{M. M\'endez et al.}
\righthead{Kilohertz QPOs in 4U 1636--53}

\begin{document}

\title{Difference Frequency of Kilohertz QPOs Not Equal to Half the
Burst Oscillation Frequency in 4U\,1636--53}

\author{Mariano M\'endez\altaffilmark{1,2},
        Michiel van der Klis\altaffilmark{1},
        Jan van Paradijs\altaffilmark{1,3},
}

\altaffiltext{1}{Astronomical Institute ``Anton Pannekoek'',
       University of Amsterdam and Center for High-Energy Astrophysics,
       Kruislaan 403, NL-1098 SJ Amsterdam, the Netherlands}

\altaffiltext{2}{Facultad de Ciencias Astron\'omicas y Geof\'{\i}sicas, 
       Universidad Nacional de La Plata, Paseo del Bosque S/N, 
       1900 La Plata, Argentina}

\altaffiltext{3}{Physics Department, University of Alabama in Huntsville,
       Huntsville, AL 35899, USA}

\begin{abstract}

We have analyzed data obtained during two observations with the Rossi
X-ray Timing Explorer on January 5 and 8, 1997, of the low-mass X-ray
binary (LMXB) and atoll source 4U 1636--53. We measure the frequency
separation of the two simultaneous kilohertz quasi-periodic
oscillations (kHz QPOs) in this source to be 253.7$\pm$4.7 and
246.4$\pm$5.4 Hz, respectively. These values are inconsistent with
being equal to 0.5 times the frequency of the 581-Hz oscillations that
have been detected previously in 4U 1636--53 during type I bursts. The
weighted average discrepancy is 39.5$\pm$3.5 Hz. This result shows
that a simple beat-frequency interpretation of the kHz QPOs, in which
the frequency of the oscillations detected during type I bursts equals
the separation between the two kHz QPOs (or twice that value), is
incorrect.

\end{abstract}


\section{Introduction}
Observations with the Rossi X-ray Timing Explorer (RXTE) have so far
revealed coherent oscillations during type I X-ray bursts in 6
low-mass X-ray binaries (LMXBs): in 4U 1728--34 at 363 Hz (Strohmayer
et al. \cite{strohmayer96a}, \cite{strohmayer97b}), in 4U 1636--53 at
581 Hz (Zhang et al. \cite{zhang96}, Strohmayer et al.
\cite{strohmayer98}), in KS 1731--26 at 524 Hz (Smith et al.
\cite{smith97}), in 4U 1702--43 at 330 Hz (Strohmayer, Swank and Zhang
\cite{ssz98}), in Aql X--1 at 549 Hz (Zhang et al. \cite{zhang98}),
and in a source near the Galactic Center (probably MXB 1743--29) at 589
Hz (Strohmayer et al. \cite{strohmayer97a}).  The first four sources
have also shown two simultaneous kilohertz quasi-periodic oscillations
(kHz QPOs) in their persistent emission (Strohmayer et al.
\cite{strohmayer96a}; Wijnands et al. \cite{wijnandsetal97}; Wijnands
\& van der Klis \cite{wijnands97}; Swank \cite{swank98}).  In 4U
1728--34 the separation between the two simultaneous kHz QPOs was $355
\pm 5$ Hz, consistent with the frequency of the oscillations during
the type I bursts. In 4U 1702--43 the kHz QPO separation frequency and
the burst oscillation frequency were also similar. In 4U 1636--53 and
KS 1731--26 the frequency separation of the kHz peaks was $ 276 \pm
10$ Hz and $260\pm 10$ Hz, respectively, consistent with half the
frequency during the burst oscillations.

These results strongly suggested that the burst oscillations and the
twin kHz QPO are connected through a beat frequency relation, in which
signals at two of the frequencies are interacting to produce a third
one at their difference frequency. There are good arguments that the
burst oscillations are due to short-lived thermonuclear-powered hot
spots on the neutron star surface that spin around with approximately
the star's spin frequency (Strohmayer et al. \cite{strohmayer98}).  In
a beat frequency interpretation a natural choice for the two
interacting oscillations then is (i) the burst oscillations, occurring
at the neutron star spin frequency (or twice that), and the (ii) the
higher-frequency kHz QPO, occurring at the Kepler frequency
corresponding to some preferred radius in the accretion disk
(Strohmayer et al. \cite{strohmayer96b}; Miller, Lamb, \& Psaltis
\cite{miller98}). The lower-frequency kHz QPO is then due to the beat
between these two frequencies. In such an interpretation the kHz QPO
frequency difference is equal to the neutron star spin frequency and
hence predicted to be constant. In most of the sources that showed the
simultaneous kHz QPO peaks the frequency difference was indeed
consistent with being constant as the twin kHz peaks moved up and down
in frequency as a function of, presumably, accretion rate.

In sources for which only the twin kHz QPO, and no burst oscillations,
were observed the frequency difference was interpreted in terms of the
neutron star spin frequency as well. However, we now know that two of
these sources do not fit this interpretation: both in Sco X-1 (van der
Klis et al. \cite{vanderklisetal97a}), and in 4U 1608--52 (M\'endez et
al. \cite{mendez98b}) the separation between the two simultaneous kHz
QPOs varies significantly, by $\sim 40$\,\% on time scales of a few
days, showing that at least in these sources the separation frequency
can not be simply the spin frequency.

In this Letter we present new results that conclusively show that the
frequency separation of the kHz QPOs in 4U 1636--53 differs from being
equal to half the frequency of the oscillations detected during type I
X-ray bursts in this source. In Section 2 we describe the observations
and data analysis. In Section 3 we discuss our findings.

\section{Observations and Data Analysis}

We have analyzed two observations of 4U 1636--53 carried out with the
Proportional Counter Array (PCA) on board RXTE during January 1997.
The first started on 1997 January 5, 22:05 UTC and lasted for $\sim
19.5$ ks, and the second one started on January 8, 04:33 UTC and
lasted $\sim 12.5$ ks.  In both observations data were collected using
an Event mode with 1/8192 s time resolution and 64 energy channels
covering the nominal 2--60 keV energy band of the PCA.  No type I
X-ray bursts were detected.

We divided these high-time resolution 2--60 keV data into segments of
64 s, Fourier transformed these, and for each segment produced a power
spectrum extending from 1/64 to 2048 Hz.  In both observations a
strong kHz QPO peak was visible.  The centroid frequency of this peak
varied between 830 Hz and 930 Hz on January 5, and between 890 Hz and
1050 Hz on January 8.  In both cases, a second, less significant peak
was detected at a frequency 200 to 300 Hz higher than that.

We repeated the procedure using data in several different energy
bands. While the lower-frequency QPO was significant in all of these
bands, the higher-frequency QPO was more significant when we only used
data above 5 keV.  For this reason, and to reduce the influence of the
background at high energies, we decided to use only the data between 5
and 20 keV. Again, we produced a power spectrum every 64 s. As the
lower-frequency QPO was well detected in each segment, we fitted its
central frequency and then shifted the frequency scale of each spectrum
to a frame of reference where the position of the most significant peak
was constant in time (see M\'endez et al. \cite{mendez98a}). Finally,
we averaged these shifted spectra to produce one single 5--20 keV power
spectrum per observation.

\begin{figure}[ht]
\plotfiddle{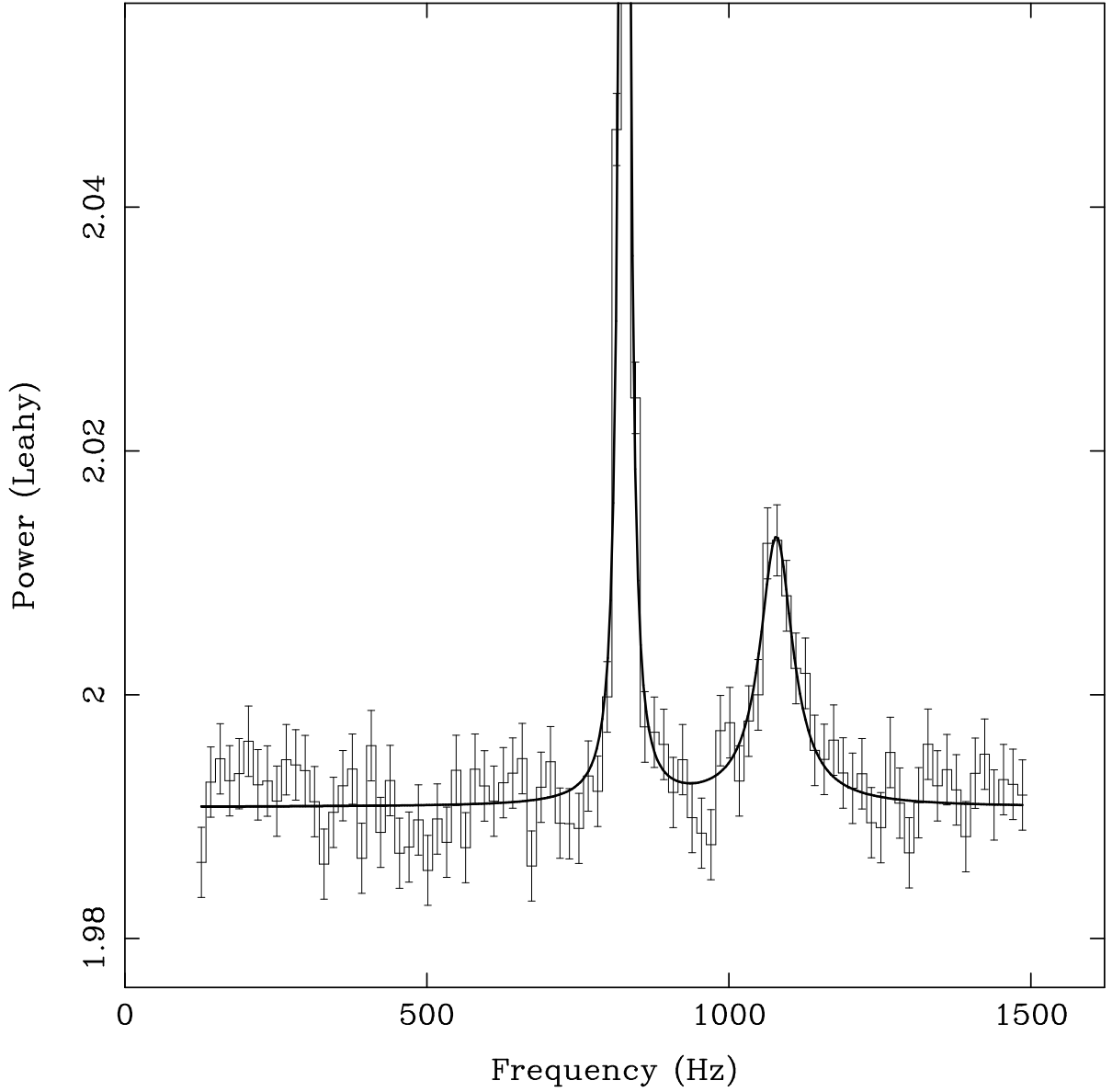}{220pt}{0}{90}{90}{-240}{-330}
\vspace{7.4cm}
\caption{
Average power spectrum of the two observations presented in this
paper, in the 5--20 keV energy range. The frequency of the
lower-frequency peak as measured in 64-s data segments was arbitrarily
shifted to improve the detection of the QPO at higher frequencies (see
M\'endez et al. \cite{mendez98a} for details). The solid line
represents the best fit model as described in the text.  The frequency
separation of the two peaks is $250.6 \pm 3.5$ Hz.
\label{figps}
}
\end{figure}

We fitted the 256--1500 Hz frequency range of each of these two average
power spectra using a function consisting of a constant, representing
the Poisson noise, and two Lorentzians, representing the QPOs.  The fits
were good, with reduced $\chi^{2} \leq 1.1$.  We show in Figure
\ref{figps} the average power spectrum of the two observations and the
model that we used to fit it.  The 5--20 keV amplitudes and FWHM of
the lower frequency QPO were $8.6 \pm 0.1$\,\% rms and $13.2 \pm 0.4$
Hz, respectively on January 5, and $3.5 \pm 0.2$\,\% rms and $6.2 \pm
0.7$ Hz, respectively on January 8.  The amplitudes
and FWHM of the higher frequency QPO were $5.2 \pm 0.4$\,\% rms and $70
\pm 12$ Hz, and $4.8 \pm 0.4$\,\% rms and $64 \pm 14$ Hz, respectively.
The frequency difference between the two simultaneous kHz QPOs was
consistent with being constant. The separation between the two peaks
was $253.7 \pm 4.7$ Hz on January 5, and $246.4 \pm 5.4$ on January
8. The average separation during these two observations as measured
from the overall average power spectrum was the same as the weighted
average of these two values: $250.6 \pm 3.5$ Hz.

\section{Discussion}

Our measured average kHz QPO peak separation, $250.6 \pm 3.5$ Hz, is
inconsistent with being equal to half the burst oscillation
frequency at better than 10$\sigma$ confidence. Although the burst
oscillation frequency shows a small ($<$2 Hz) drift during some
bursts, it is statistically very well determined and seems to always
be between 579 and 582 Hz (see Strohmayer, Swank and Zhang
\cite{ssz98}, and Strohmayer et al. \cite{strohmayer98}). So, the
separation frequency is too small by between 38.9$\pm$3.5 and
40.4$\pm$3.5 Hz to be equal to half the burst oscillation frequency.

In the beat frequency model for the kHz QPOs (Section 1), the two
basic frequencies of the system are the Keplerian frequency of the
accreting material in orbit around the neutron star at some preferred
radius, identified with the high frequency kHz QPO, at a frequency
$\nu_{\rm upp}$, and the spin frequency of the neutron star, $\nu_{\rm
s}$.  These two frequencies beat with each other to produce the
low-frequency kHz QPO at $\nu_{\rm low} = \nu_{\rm upp} - \nu_{\rm
s}$. So, the model predicts that $\Delta\nu\equiv\nu_{\rm upp}-\nu_{\rm
low}=\nu_s$.

As noted in Section 1, this model agrees well with the results
obtained in 4U 1728--34 (Strohmayer et al.  \cite{strohmayer96a}) if
one assumes that $\nu_{\rm s}$ in this source is equal to the
frequency $\nu_{\rm burst}$ of the oscillations observed during
bursts.  In 4U 1636--53 and KS 1731--26, $\Delta \nu$ is approximately
equal to $\nu_{\rm burst}/2$ (Wijnands et al. \cite{wijnandsetal97};
Wijnands \& van der Klis \cite{wijnands97}).  A beat-frequency model
could perhaps account for this if in these two sources $\nu_{\rm
s}=\nu_{\rm burst}/2$. This could occur if, for example, two hot spots
are present during the burst. (This interpretation has been contested
by Strohmayer et al. \cite{strohmayer98} on the basis of the large
observed amplitudes of the oscillations.)

Our results show conclusively that, at least in 4U 1636--53, this
simple beat frequency picture is invalid.  If we assume that $\nu_{\rm
S} = \nu_{\rm burst}/2$, then $\nu_{\rm s}$ exceeds $\Delta\nu$ by a
value near 40$\pm$3.5\,Hz, a discrepancy of more than $10\sigma$. (Of
course, the situation would be worse if $\nu_{\rm s}=\nu_{\rm
burst}=581$ Hz).

The beat-frequency model for the twin kHz QPO in LMXBs has been
strongly challenged by the measurements of a variable peak separation
in Sco X-1 and 4U 1608--52 (van der Klis et
al. \cite{vanderklisetal97a}; M\'endez et al.  \cite{mendez98b}). Our
findings show that the beat frequency model in its current form fails
in the case of 4U\,1636--53.  This is the first time that this is
demonstrated in an object where all three of the frequencies involved
in the beat-frequency interpretation are measurable.  In view of the
great similarity of the kHz QPO phenomenology in all LMXBs this
conclusion also applies to the phenomenon as observed in other LMXBs.

It may be too early to discard the beat frequency model altogether. It
is conceivable that the kHz peak separation frequency, while not being
equal to the neutron star spin frequency (or twice that) is still
constrained to be relatively near it. This could be the case if it is
the spin frequency of accreting matter that is not corotating with the
star. It is also possible that the beat frequency is generated by
processes occurring at a radius that differs from the radius inferred
from the Kepler frequency interpretation of the upper kHz QPO
frequency. In view of the complexity apparently involved in a
successful beat frequency model for the kHz QPO, in our opinion
constraints on physical parameters such as neutron star spin frequency,
mass, and radius inferred from the properties of these QPO, must for
the time being be interpreted with great care.

This work was supported in part by the Netherlands Organization for
Scientific Research (NWO) under grant PGS 78-277 and by the Netherlands
Foundation for research in astronomy (ASTRON) under grant 781-76-017.
MM is a fellow of the Consejo Nacional de Investigaciones
Cient\'{\i}ficas y T\'ecnicas de la Rep\'ublica Argentina.  JVP
acknowledges support from the National Aeronautics and Space
Administration through contract NAG 5-3269 and 5-4482.

\onecolumn
\clearpage

\end{document}